\documentstyle[prl,aps]{revtex}
\begin{document}
\draft \title{Surface superconductivity and order parameter
suppression in UPt$_3$ } \author{D.F. Agterberg and M.B. Walker}

\address{Department of Physics, University of Toronto, Toronto, Ontario,
Canada, M5S 1A7}
\date{\today}
\maketitle
\begin{abstract}
We show that a recent measurement of surface superconductivity in
UPt$_3$ (Keller {\it et. al.}, Phys. Rev. Lett. {\bf 73}, 2364 (1994))
can be understood if the superconducting pair wavefunction is
suppressed anisotropically at a vacuum to superconductor
interface.  Further measurements of surface superconductivity can
distinguish between the various phenomenological models of
superconducting UPt$_3$.
\end{abstract}

Superconductivity in hexagonal UPt$_3$ exhibits many properties unique
among superconductors. The most striking of these is the pressure
temperature magnetic field phase diagram that shows several
superconducting phases \cite{ade90,dij93,bou94}. For example, a graph
or the upper critical field for fields in the basal plane
($H_{c_2}^{ab}$) versus temperature exhibits a kink at a temperature
$T^*$ \cite{has89}. This kink can be explained in terms of a
generalized Ginzburg-Landau model in which the order parameter has two
complex components, $\eta_1$ and $\eta_2$
\cite{mac89,hes89,joy90,sau94}.  For temperatures below $T^*$, one of
these components orders at $H_{c_2}^{ab}$, whereas for temperatures
above $T^*$ the other component orders at $H_{c_2}^{ab}$.

Recently, a detailed study of surface superconductivity in UPt$_3$ by
Keller {\it et. al.}  has produced puzzling features not observed in
other superconductors \cite{kel94,kelt} (see also \cite{tai90}).  In
particular, the ratio of the upper critical field for surface
superconductivity ($H_{c_3}$) to that for bulk superconductivity has
an unusual temperature dependence \cite{kelt}.  Furthermore, different
surfaces of the UPt$_3$ whisker were observed to have different values
of $H_{c_3}$ \cite{kel94,kelt}.  It is argued here that the reason for
this unusual temperature dependence is that one of the components for
the order parameter (say $\eta_1$) is suppressed at the surface (and
hence cannot sustain surface superconductivity); the transition at
$H_{c_3}$ is thus to a superconducting $\eta_2$ state, both below and
above $T^*$. Order parameter suppression at a superconductor to vacuum
interface is known not to occur in conventional isotropic
superconductors \cite{stj69} but is expected on theoretical grounds
(although not yet observed experimentally) for some unconventional
superconductor symmetries \cite{sig91,sam95}. We believe the observed
behavior of $H_{c_3}/H_{c_2}$ in UPt$_3$ is the only evidence to date
of an anisotropic suppression of the superconducting order parameter
at a vacuum to superconductor interface. This interpretation of the
measurements of Keller {\it et. al.} imposes strong constraints on
existing phenomenological models of superconductivity in UPt$_3$ and
indicates that such measurements can provide a new technique for
determining the symmetry of anisotropic superconductors.

In this article we will initially consider surface superconductivity
for the $E$-REP Ginzburg-Landau models proposed by by Machida and
Ozaki \cite{mac89} and by Hess {\it et. al.} \cite{hes89} and improved
by Sauls \cite{sau94} (an important element of these models is the
hexagonal symmetry breaking due to the basal plane antiferromagnetism
\cite{aep89,hay92}). We then show how the inclusion of symmetry
dependent surface contributions to the free energy is essential for
understanding the experiments.  Finally, we show how surface
superconductivity studies can distinguish between these models, the
1D-REP model \cite{mac91}, and the $AB$ model \cite{che93}.

For the odd parity $E_{2u}$ and $E_{1u}$ models the gap function is
described by the pseudo-spin pairing gap matrix
$\Delta(\vec{r},\vec{k})=i[\eta_1(\vec{r})\vec{d_1}(\vec{k})\cdot\vec{\sigma}+
\eta_2(\vec{r})\vec{d_2}(\vec{k})\cdot\vec{\sigma}]\sigma_y$ where
$\sigma_i$ are the Pauli matrices. We will assume that
$\vec{d}_i(\vec{k})=d_i(\vec{k})\hat{z}$; this form of the gap
function is required in the $E_{2u}$ model to account for the gap
nodal structure implied by experiment \cite{sau94}.  For the even
parity $E_{2g}$ and $E_{1g}$ models the form of the gap matrix is
$\Delta (\vec{r},\vec{k})=i\sigma_y[\eta_1(\vec{r})\psi_1(\vec{k})+
\eta_2(\vec{r})\psi_2(\vec{k})]$. The free energy density for these
$E$-REP models is
\begin{eqnarray}
F= && \alpha_0 (T-T_c) \vec{\eta} \cdot \vec{\eta}^* -\gamma
(|\eta_2|^2-|\eta_1|^2)+\kappa_1^+|D_i\eta_2|^2+\kappa_1^-|D_i\eta_1|^2\nonumber
\\ && \kappa_4^+|D_z\eta_2|^2+ \kappa_4^-|D_z\eta_1|^2 + \kappa_2
(D_i\eta_i)(D_j\eta_j)^* + \kappa_3 (D_i\eta_j)(D_j \eta_i)^* +{{\vec{h}^2}/{8
\pi}} -{{\vec{h}\cdot \vec{H}}/{4 \pi}}
\label{f3}
\end{eqnarray}
where $\vec{D}=\vec{\partial}-(i2e/\hbar c)\vec{A}$,
$\vec{h}=\vec{\partial}\times\vec{A}$, $\vec{H}$ is the applied field,
$i$ and $j$ refer to indices $x$ (and 1) and $y$ (and 2), and
$\kappa_l^{\pm}=\kappa_l(1 \pm \delta_l)$.  The convention
$\sigma_{\vec{a}}\eta_1=-\eta_1$ for even parity representations and
$\sigma_{\vec{a}}\eta_1=\eta_1$ for the odd parity representations
fixes the axes and $\sigma_{\vec{a}}$ is a reflection in a plane
normal to the $\vec{a}$ crystallographic axis (these conventions
allow the discussion of surface superconductivity to occur on the
same footing for the even and odd parity representations). For the
calculation of $H_{c_2}$ and $H_{c_3}$ higher order terms in $\eta_1$
and $\eta_2$ in Eq.~\ref{f3} can be omitted.  The coefficients
$\gamma$ and $\delta_l$ arise from the coupling to the
antiferromagnetic moment. The gradient terms with coefficients
$\kappa_2$ and $\kappa_3$ give rise to anisotropy in $H_{c_2}^{ab}$
\cite{mac89,hes89,agt951}. This has been observed to be of the order
$0.01 H_{c_2}^{ab}$ \cite{kel94}. It has also been argued by Sauls
\cite{sau94} that these terms are small relative to $\kappa_1$ in the
$E_{2u}$ model. We will therefore solve for $H_{c_3}$ to first order
in $\lambda=(\kappa_{2}+\kappa_3)/2\kappa_1$. The linearized
Ginzburg-Landau equation given by the free energy in Eq.~\ref{f3} is

\begin{equation}
\alpha \vec{\eta} = \pmatrix{\kappa_{1^-23}D_x^2 +
\kappa_1^-D_y^2+\kappa_4^-D_z^2 -\gamma & \kappa_2 D_xD_y +
\kappa_3 D_y D_x \cr \kappa_3 D_x D_y + \kappa_2 D_y D_x &
\kappa_{1^+23}D_y^2 + \kappa_1^+D^2_x+\kappa_4^+D_z^2+\gamma  \cr}
\vec{\eta}
\label{GL}
\end{equation}
where $\kappa_{1^\pm 23}=\kappa_1^{\pm}+\kappa_2+\kappa_3$.

We will consider the geometry in which the superconductor occupies the
half space $y>0$. The surface normal will be taken to be along the
$\vec{a}$ (or $y$) direction and the antiferromagnetic moment along
the $\vec{a}^*$ (or $x$) direction.  The field will be chosen to lie
along the $\vec{a}$ or $\vec{a}^*$ directions. This geometry will
allow for a comparison to be made with the measurements of Keller {\it
et. al.}  \cite{kel94,kelt}.

The off diagonal terms in Eq.~\ref{GL} containing $\kappa_2$ and
$\kappa_3$ vanish in the determination of the upper critical field for
the field orientations chosen above when there is no surface
present. In the presence of a surface normal along $\vec{a}$ it can be
shown that to first order in $\lambda$ these terms have no effect.  In
this limit the Ginzburg-Landau equations decouple for the $\eta_1$ and
$\eta_2$ components. When the field lies along $\vec{a}$ then the
upper critical field (denoted $H_{c_2}^a$) is given by

\begin{eqnarray}
\sqrt{\kappa_1^+\kappa_4^+}[H_{c_2}^{a}]_2=&& -\alpha+\gamma
,\hphantom{abc} \eta_2\ne 0 \hphantom{abc} \eta_1=0 \nonumber\\
\sqrt{\kappa_1^-\kappa_4^-}(1+\lambda)[H_{c_2}^{a}]_1=&&
-\alpha-\gamma,\hphantom{abc} \eta_1\ne 0\hphantom{abc} \eta_2=0.
\end{eqnarray}
where $\alpha=\alpha_0(T-T_c)$.  If $\gamma>0$ and
$\delta_1+\delta_4>0$ or $\gamma<0$ and $\delta_1+\delta_4<0$ then the
two solutions will cross at a temperature $T^*$. This crossing gives
rise to the kink in $H_{c_2}^a(T)$ that is observed in experiment
\cite{has89}. For the field along the $\vec{a}^*$ direction the upper
critical field for surface superconductivity (denoted
$H_{c_3}^{a,a^*}$ where $a$ refers to the surface normal and $a^*$ to
the field direction) can be found using the exact solution for
conventional superconductors \cite{stj69}; this yields

\begin{eqnarray}
\sqrt{\kappa_1^+\kappa_4^+}(1+\lambda)\mu^2 [H_{c_3}^{a,a^*}]_2=&&
-\alpha+\gamma,\hphantom{abc} \eta_2\ne 0 \hphantom{abc} \eta_1=0\\
\sqrt{\kappa_1^-\kappa_4^-}\mu^2 [H_{c_3}^{a,a^*}]_1=&& -\alpha-\gamma
\hphantom{abc} \eta_1 \ne 0 \hphantom{abc} \eta_2=0
\end{eqnarray}
where $\mu^2=0.59010$.  An immediate consequence is that
$H_{c_3}^{a,a^*}$ will have a kink where the two solutions cross.
This kink will occur at the approximately $T^*$ (the difference is of
order $\lambda$). Note that
$H_{c_3}^{a,a^*}/H_{c_2}^{a}=1/\mu^2+O(\lambda)$ for all temperatures
in this model. Keller {\it et. al.} \cite{kel94,kelt} have observed
that this ratio is strongly temperature dependent.  This model cannot
account for the observed behavior.

Qualitatively different behavior from that presented above arises when
the surface free energy is considered \cite{sam95}. For conventional
isotropic superconductors with order parameter $\psi$, the surface
free energy takes the form $g \int_{surface} |\psi|^2 dS$.  For $g>0$
this term suppresses surface superconductivity. However,
superconductor to insulator or vacuum boundaries of isotropic
superconductors are well described by $g=0$ \cite{stj69}. For
anisotropic superconductors this is not necessarily the case. For the
$E$-REP models and the geometry defined above the surface free energy
density has the form
\begin{equation}
F_{surface}= g_1 |\eta_1|^2 + g_2 |\eta_2|^2.
\label{none}
\end{equation}
In the presence of this surface free energy density the boundary
conditions are
\begin{eqnarray}
-(\kappa_{1}^-D_y \eta_1 +\kappa_3D_x\eta_2)=&& g_{1} \eta_1
\nonumber \\ -(\kappa_{1^+23}D_y\eta_2 + \kappa_2 D_x \eta_1)=&&
g_{2}\eta_2.
\end{eqnarray}
Assuming that the coefficients are field independent one can find
$H_{c_3}^{a,a^*}$ by using the method similar to that used for an
isotropic superconductor \cite{stj69}. For convenience we set
$\lambda=0$ in the remainder of this article (the corrections
$O(\lambda)$ are easily obtained). The solution is given by
\begin{equation}
[H_{c_3}^{a,a^*}/H_{c_2}^{a^*}]_i = {{1}\over{\mu^2(l_i)-l_i^2}}
\label{sur1}
\end{equation}
where $i$ refers to the solutions $\vec{\eta}_1=(1,0)$ or
$\vec{\eta}_2=(0,1)$, $l_1=[H^{a}_{c_2}/H_{c_3}^{a,a^*}]_1^{1/2}
g_{1}/(\kappa_1^- \alpha_0(T_c^--T))^{1/2}$,
$l_2=[H_{c_2}^{a}/H_{c_3}^{a,a^*}]_2^{1/2}
g_{2}/(\kappa_1^+\alpha_0(T_c^+-T))^{1/2}$,
$T_c^{\pm}=T_c\pm\gamma/\alpha_0$, and $\mu(l_i)$ is defined by
\begin{equation}
\int_0^\infty (2u-\mu-l_i)e^{-(u-\mu)^2}u^{-(1+\mu^2-l_i^2)/2}du =0.
\label{sur2}
\end{equation}

Insight into the values of the coefficients $g_1$ and $g_2$ can be
obtained from the microscopic model originally studied by Ambegaokar
{\it et. al.} for superfluid $^3$He \cite{amb74} and extended to
anisotropic superconductors by Sigrist and Ueda \cite{sig91} and also
by Samokhin \cite{sam95}.  The microscopic weak coupling analysis of
\cite{sig91} and \cite{sam95} indicates that for a specularly
reflecting surface with normal $\vec{n}$ $g_i=0$ if $P_{\vec{n}}
\Delta(\vec{k}) = \Delta (\vec{k})$ where $P_{\vec{n}}
\Delta(\vec{k})=\Delta(\vec{k}-2 \vec{n}\cdot \vec{k} \vec{n})$ and
that $g_i=\infty$ \cite{sam95} or $g_i\approx \alpha_0 T_c
\zeta_{0,i}$ \cite{sig91} ($\zeta_{0,i}$ is the zero temperature
coherence length) if $P_{\vec{n}} \Delta(\vec{k}) = -\Delta(\vec{k})$.
This is in agreement with experiments for conventional isotropic
superconductors for which $P_{\vec{n}} \Delta (\vec{k})=
\Delta(\vec{k})$ for all surface normals with the consequence that
$g=0$. For a surface normal along $\vec{a}$ the above considerations
imply that $g_2=0$ and $g_1=\infty$ \cite{sam95} or $g_1\approx
\alpha_0 T^-_c \zeta_{0,1}$ \cite{sig91} with $\zeta_{0,1}=
(\kappa_1^-/\alpha_0 T_c^-)^{1/2}$ \cite{note}. If we choose
$g_1=\alpha_0 T^-_c \zeta_{0,1}$ Eqs.~\ref{sur1} and \ref{sur2}
indicate that $H_{c_3}^{a,a*}$ is equal to $H_{c_2}^{a}$ to within 1/2
a percent for the temperature range we consider, an even stronger
suppression of surface superconductivity will occur for any
$g_1>\alpha_0 T^-_c \zeta_{0,1}$. Consequently $g_1\ge\alpha_0 T^-_c
\zeta_{0,1}$ is equivalent in practice to $g_1=\infty$. With these
values for the coefficients $g_i$ two distinct $H_{c_3}$ curves are
possible in UPt$_3$, depending on whether $\gamma$ and
$\delta_1+\delta_4$ are both positive or both negative. If $\gamma>0$
and $\delta_1+\delta_4>0$ then the bulk transition is to the $\eta_2$
state ({\it i.e.}  $\vec{\eta}=(0,1)$) for $T^*<T<T_c^+$ and to the
$\eta_1$ state for $0<T<T^*$.  Since the $\eta_1$ state is suppressed
at the surface, surface superconductivity can only occur in the
$\eta_2$ state and the resulting temperature dependence of
$H_{c_3}^{a,a^*}$ is shown in Fig.~\ref{fig1}~a (recall that for
$g_2=0$, $[H_{c_3}^{a,a^*}]_2=1.695 [H_{c_2}^{a}]_2$).  On the other
hand, for $\gamma<0$ and $\delta_1+\delta_4<0$, the bulk transition is
to the $\eta_1$ state for $T^*<T<T_c^+$ and to the $\eta_2$ state for
$0<T<T^*$.  In this case the behavior in Fig.~\ref{fig1}~b arises.

The above discussion is relevant to specular reflecting surfaces which
in fact does not account for the experimental results on UPt$_3$ (see
below). If diffusive scattering is considered then the $\eta_2$
component will also be suppressed at the surface. This is taken into
account by introducing $g_2\ne 0$. In Fig.~\ref{fig2} we have plotted
$H^{a,a^*}_{c_3}/H^a_{c_2}$ versus temperature for $g_1\ge\alpha_0
T^-_c \zeta_{0,1}$ and $g_2=\alpha_0 T^+_c \zeta_{0,2}/10$ where
$\zeta_{0,2}=(\kappa_1^+/\alpha_0 T^+_c)^{1/2}$.

Keller has given a graph of $\Delta
T(H)=T_{H_{c3}^{a,a^*}}(H)-T_{H_{c2}^a}(H)$ versus $H$ \cite{kelt}
that illustrates the temperature dependence of
$H_{c_3}^{a,a^*}/H_{c_2}^a$.  It was observed that $\Delta T(H)$
initially increased with increasing field until it reached a maximum
(at $H\approx 0.25$ Tesla) after which it {\it decreased} with
increasing field. It can be seen that for specular reflection $\Delta
T(H)$ is always increasing with $H$ (see Fig.~1a). The experimental
behavior is best described by $\gamma>0$, $\delta_1+\delta_4>0$, and
the diffusive scattering values $\alpha_0 T^+_c \zeta_{0,2}/12 <g_2
<\alpha_0 T^+_c \zeta_{0,2}/8$ and $g_1\ge \alpha_0 T^-_c \zeta_{0,1}$
(as in Fig.~2).

We have considered the simplest case in which the antiferromagnetic
moment is pinned orthogonal to the surface.  Other possible relative
orientations of the surface and the antiferromagnetic moment are
possible. Differing domain distributions at different surfaces may
account for the different values of $H_{c_3}$ observed at these
surfaces (note that different values of $g_2$ for different surfaces
also can account for the observed $H_{c_3}$ values). A detailed study
can be undertaken once a better understanding of the antiferromagnetic
domain structure near the surface is achieved.  We have also assumed
that the whisker used by Keller {\it et. al.} can be described by the
same model that describes annealed UPt$_3$ samples.  We note that
Keller gives the $H_{c_2}^a$ curve for this whisker \cite{kelt}.
There is a strong curvature in $H_{c_2}^a$ at small fields and as the
field increase above $H\approx 0.25$ Tesla (note this is approximately
the same field at which $\Delta T(H)$ attains its maximum),
$H_{c_2}^a$ increases linearly with temperature. We expect that our
model is qualitatively correct for this whisker.  However, experiments
on UPt$_3$ samples with a well defined kink in $H_{c_2}^a$ will be
required for a quantitative comparison.

The model presented above also implies informative results for
$H_{c_3}^{c,a}$. For a surface normal along the $\vec{c}$ axis
symmetry implies $g_1=g_2$ in Eq.~\ref{none}. It is therefore expected
that $H_{c_3}^{c,a}$ will either exhibit a kink at $T\approx T^*$ or
be suppressed. Using the microscopic arguments presented earlier it is
expected that for the $E_{2u}$ and $E_{1g}$ representations
$H_{c_3}^{c,a}\approx H_{c_2}^{a}$ while for the $E_{1u}$ and $E_{2g}$
representations $H_{c_3}^{c,a}>H_{c_2}^{a}$.

A similar analysis for different models of UPt$_3$ results in
qualitatively different behavior than that of the $E$-REP models. In
the 1D-REP model \cite{mac91} the order parameter is a three
dimensional order parameter that transforms as a vector under spin
rotations and as a one dimensional representation under spacial
transformations. In this model the antiferromagnetic moment breaks the
spin degeneracy of the order parameter.  Since the order parameter
transforms as a one dimensional representation under spacial
transformations the surface free energy density takes the form $g
|\vec{\eta}|^2$ for all surface normals (note that this model is
exactly solvable for $H_{c_3}^{\vec{n},a^*}$ using Eqs.~\ref{sur1} and
\ref{sur2}).  This surface free energy does not allow any anisotropy
to occur between the different order parameter components. It is
therefore difficult to reconcile this model with the experimental
observations of Keller {\it et.al.}  \cite{kel94,kelt}.

In the accidentally nearly degenerate $AB$ model \cite{che93} the
order parameter has two components of the same parity that transform
as different one dimensional representations of the hexagonal point
group (one as an $A$ and the other as a $B$ representation).  The
surface free energy for all surface normals takes the form $g_A
|\psi_A|^2 + g_B |\psi_B|^2$.  This model is also exactly solvable for
$H_{c_3}$. Analysis indicates that the results for $H_{c_3}^{a,a^*}$
are similar to that of the $E$-REP models with $\lambda=0$. However, a
qualitative difference may occur in $H_{c_3}^{c,a}$ between these two
models. For $H_{c_3}^{c,a}$ in the even parity $AB$ models,
microscopic analysis indicates that one component will be suppressed
and the other will not be.  Consequently the resulting behavior of
$H_{c_3}^{c,a}$ will correspond to that of one of the two curves in
Fig.~\ref{fig2}.

In conclusion, we have, for the first time, used experimental results
on surface superconductivity to yield information about the symmetry
of the superconducting order parameter.  In particular the experiments
of Keller {\it et. al.}  \cite{kel94,kelt} on UPt$_3$ can be
understood if one component of the order parameter is strongly
suppressed while the other component is weakly suppressed for a
surface normal along the $\vec{a}$ direction.  This anisotropy places
a strong constraint on phenomenological models of superconductivity in
UPt$_3$. Further experimental investigations of surface
superconductivity can place additional constraints on these models.

We acknowledge the support of the Natural Sciences and Engineering
Research Council of Canada. D.F.A. was also supported by the Walter
C. Sumner Foundation. We also wish to thank Niels Keller for providing
us with a copy of his thesis.

\figure{\caption{$H_{c_3}^{a,a^*}$ (upper curves) and $H_{c_2}^a$
(lower curves) as a function of temperature for different model
parameters for specular reflecting surfaces.  The parameters have been
chosen to agree with the experimental phase diagram for
$H_{c_2}^a$. a) $\gamma>0$, $\delta_1+\delta_4>0$ b) $\gamma<0$,
$\delta_1 +\delta_4 <0$. \label{fig1}}

\figure{\caption{The ratio $H_{c_3}^{a,a^*}/H_{c_2}^a$ for diffusive
scattering at the surface. The rightmost curve corresponds to
$\gamma>0$, $\delta_1+\delta_4>0$ and the other curve to $\gamma<0$,
$\delta_1+\delta_4<0$. \label{fig2}}


\begin{references}
\bibitem{note} This conclusion requires $\vec{d}_i= \hat{z} d_i$ for
the odd parity representations. If other spin components are included
then $g_2\ne 0$.
\end{references}
\end{document}